\documentclass[aps,prb,showpacs]{revtex4}
\usepackage{graphicx, amsmath, amsfonts, amssymb}

\begin{document}

\title{Two-phonon scattering in graphene in the quantum Hall regime}

\author{A. M. Alexeev}
\affiliation{School of Physics, University of Exeter, Stocker Road, Exeter EX4 4QL, United Kingdom}

\author{R. R. Hartmann}
\affiliation{De La Salle University-Manila, 2401 Taft Avenue, 1004 Manila, Philippines}

\author{M. E. Portnoi}
\email[]{m.e.portnoi@exeter.ac.uk}
\affiliation{School of Physics, University of Exeter, Stocker Road, Exeter EX4 4QL, United Kingdom}
\affiliation{International Institute of Physics, Universidade Federal do Rio Grande do Norte, Natal, RN, Brazil}

\date{October 31, 2015}

\begin{abstract}
One of the most distinctive features of graphene is its huge inter-Landau-level splitting in experimentally attainable magnetic fields which results in the room-temperature quantum Hall effect. In this paper we calculate the longitudinal conductivity induced by two-phonon scattering in graphene in a quantizing magnetic field at elevated temperatures. It is concluded that the purely phonon-induced scattering, negligible for conventional semiconductor heterostructures under quantum Hall conditions, becomes comparable to the disorder-induced contribution to the dissipative conductivity of graphene in the quantum Hall regime.
\end{abstract}


\maketitle

\section{Introduction}

\label{Inroduction}

The quantum Hall effect, discovered in 1980 by Klaus von Klitzing,\cite{Klitzing1980} allows one to determine the quantum resistance standard in terms of the electron charge and Plank's constant with a spectacular accuracy. However, the level of precision necessary for metrology applications (a few parts per billion) requires for conventional quasi-two-dimensional semiconductor systems the use of ultra-low temperatures and high magnetic fields.\cite{Jeckelmann1997, Jeckelmann2001}
The discovery of graphene\cite{Novoselov2004} in 2004 led to a revival of interest in quantum Hall effect physics.\cite{Novoselov2005, Zhang2005} The energy of Landau Levels (LLs) in graphene is given by
\begin{equation}
\label{LandauLevels}
E_{N}= \pm (\hbar v_{\mathrm{F}} / l_{B}) \sqrt{2N} \mbox{,}
\end{equation}
where $v_{\mathrm{F}} = 1 \times 10^6 \mbox{ } \mathrm{m/s}$ is the Fermi velocity, $l_{B}=\sqrt {\hbar / eB}$ is the magnetic length, $B$ is the magnitude of the applied magnetic field, $N$ is the LL number, and the ``$\pm$'' signs correspond to the electrons and holes respectively. \cite{McClure1956, Geim2007, Neto2009} For $B= 10~\mathrm{T}$ the separation between the zero and the first LL in graphene is $\Delta E_{01} / k_{\mathrm{B}}\approx1300~\mathrm{K}$ (in comparison, $\Delta E / k_{\mathrm{B}} \approx 200 \mbox{ } \mathrm{K}$ in conventional quasi-two-dimensional semiconductor systems). Therefore, the quantum Hall effect in graphene can be observed even at room temperatures in experimentally attainable magnetic fields. Indeed, the room-temperature quantum Hall effect in graphene was first observed in 2007;\cite{Novoselov2007} however, extremely high magnetic fields (up to $45~\mathrm{T}$) were needed.

It has been understood since the 1930's that the longitudinal conductivity of metals in a quantizing magnetic field increases with increasing electron scattering.\cite{Titeika1935} To develop a graphene-based quantum Hall standard of resistance,\cite{
Giesbers2008, Tzalenchuk2010, Poirier2010, Webber2013} which would work at elevated temperatures and in moderate magnetic fields, it is essential to examine the contributions from different scattering processes to the longitudinal conductivity $\sigma_{xx}$ in the quantum Hall regime (QHR), $T \ll \Delta E_{01}/ k_{\mathrm{B}}$, as $\sigma_{xx}$ provides the major correction to the quantized value of the Hall resistance. In conventional semiconductor systems the quantum Hall effect is observed at liquid helium temperatures only, and the value of $\sigma_{xx}$ is governed by scattering on disorder. In the QHR, $\sigma_{xx}$ depends exponentially on the energy separation between the Fermi level and the nearest LL. The longitudinal conductivity prefactor has a theoretically-predicted value of $e^{2}/h$ in the presence of short-range disorder and $2e^{2}/h$ in the presence of long-range disorder.\cite{
Polyakov1994, Polyakov1995} It was shown that the contribution from phonon scattering is negligibly small. The exceptions are exotic cases such as magnetoroton dissociation, in which the whole effect arises from scattering on phonons,\cite{Apalkov2002} or an enhancement of phonon-induced scattering near the intersection of two LLs corresponding to different size-quantization subbands.\cite{Apalkov2002TwoLevelPRB, Apalkov2002TwoLevelPhE,Golovach2006} However, as we show in this paper, the phonon-scattering mechanism dominates in the high-temperature QHR in graphene, since at $T > T_{l_{B}}= (\hbar s / l_{B})/k_{\mathrm{B}}$ (where $s=2 \times 10^{4}~\mathrm{m/s}$ is the sound velocity in graphene \cite{Ono1983, Klemens1994}) the energy of an acoustic phonon with a wavevector comparable to the inverse magnetic length is much smaller than the temperature; therefore, the number of such phonons increases drastically.

In this paper we restrict our consideration to electron scattering induced by the interaction with intrinsic in-plane phonons only, neglecting the effects of disorder as well as electron-electron interactions. Our calculations result in the lower estimate of the longitudinal conductivity $\sigma_{xx}$ in graphene in the QHR, as we do not study electron interactions with various other types of phonons such as out-of-plane flexural phonons, which are present in suspended samples,\cite{Mariani2008, Castro2010} or bulk acoustic phonons in a substrate, which interact with electrons in graphene on polar substrates such as boron nitride \cite{Wang2013} or silicon carbide via piezoelectric coupling.\cite{Geick1966, Hwang1978}

\section{Theoretical formalism: Two-phonon scattering within one Landau level}

\label{Theory}

The wave function of a charged carrier (electron or hole) in graphene subjected to a magnetic field $\mathbf{B}$ normal to the graphene plane is given in the Landau gauge, $\mathbf{A}=\left (0,Bx,0\right )$, by the following expression:\cite{McClure1956, Geim2007, Neto2009}
\begin{equation}
\label{2componentWF}
\Psi_{N, k_{y}}=  \frac {C_{N}}{\sqrt{2^N N! \sqrt{\pi} l_{B}}} \exp \left[- \frac {\left(x-x_{0} \right)^{2}}{2 l_{B}^{2}} \right] \left[{H_{N}\left[ \left(x-x_{0}\right) / l_{B} \right] \atop \pm \sqrt{2 N} H_{N-1} \left[ \left(x-x_{0}\right) / l_{B} \right]}\right] \frac {\exp \left(ik_{y} y\right)} {\sqrt{L_{y}}}
\mbox{,}
\end{equation}
where $C_{0}=1$ and $C_{N \neq 0}=1/\sqrt{2}$, $x_{0} = l_{B}^{2} k_{y}$ is the guiding centre coordinate, $k_{y}$ is the electron wavevector $y$-component, $H_{N}$ are the Hermite polynomials, $L_{x}$ and $L_{y}$ are the graphene sample dimensions, the ``$\pm$'' sign corresponds to electrons and holes, and the spin and valley indeces are omitted. Here the $z$-axis has been chosen in the direction of the applied magnetic field and the $x$-axis and $y$-axis are in the graphene plane. The wave function given by Eq.\,(\ref{2componentWF}) is defined for one Dirac point only and its two components correspond to graphene's two sublattices. In what follows all results will be obtained for electrons only. Calculations for the non-equivalent Dirac point and holes can be easily repeated in a similar fashion. From here on we will omit the ``$\pm$'' sign in order to simplify notation. We account for the valley and spin degeneracy by multiplying the final result for $\sigma_{xx}$ by the factor of four at the final stage of calculations.

Electron scattering on phonons leads to a change in the $k_{y}$ component of the electron wavevector, which results in a change to the electron guiding centre coordinate  $x_{0}$. Electron transitions between LLs due to one phonon scattering are suppressed due to the large energy gaps between LLs in graphene in the QHR. Indeed, the number of phonons with the energy required for such transitions is
$
n_{q} \simeq \exp \left ( - \Delta E_{N_{1}N_{2}} /k_{\mathrm{B}} T \right )
$
($\Delta E_{N_{1}N_{2}}$ is the energy gap between two different LLs in graphene), which is very small in the QHR. Inter-LL scattering on acoustic phonons has an additional exponentially strong suppression in the matrix element of transition which is markedly different from the case of conventional semiconductor systems as discussed in the Appendix. It is also evident that in the QHR only scattering on acoustic phonons can provide a noticeable contribution to the longitudinal conductivity $\sigma_{xx}$. The optical phonon energy in graphene corresponds to the temperature range $1800~\mathrm{K}-2300~\mathrm{K}$~\cite{
Lazzeri2006, Suzuura2008} which leads to very small optical phonon occupation numbers at room temperatures and below. In this work, we are interested in phonon-induced equilibrium longitudinal conductivity, whereas, magneto-phonon resonance associated with optical phonons has been studied in Ref. \cite{Goerbig2007, Mori2011, Kim2013}. One-phonon scattering within the same LL is ineffective because of the small width of the LLs, which can be arguably achieved in pristine graphene samples. Interestingly, even in heavily disordered samples the broadening of LLs was found to shrink in several particular cases. \cite{Giesbers2007, Pereira2011} For our calculations we assume an infinitely narrow band of delocalized states in the middle of each LL with a vanishing density of states in between LLs. In this limit one-phonon scattering within one LL is forbidden by energy and momentum conservation. The next order process to consider is two-phonon scattering through virtual states with no change in the electron initial and final LL numbers but with the change in its in-plane momentum (or the guiding centre coordinate). We calculate the longitudinal conductivity $\sigma_{xx}$ at the $N$th LL due to two-phonon scattering using the generalized form of the Einstein relation:\cite{Bychkov1981, Dmitriev1995, Golovach2006}
\begin{equation}
\label{LongConductivity}
\sigma_{xx} = \frac {e^2} {2 \pi l_{B}^2} \nu_{N} \left ( 1- \nu_{N} \right ) \frac {D} {k_{\mathrm{B}} T} \mbox{,}
\end{equation}

\begin{equation}
\label{DiffCoefficient}
D = \frac {l_{B}^4} {2} \sum_{k_{y}'} W_{ k_{y} \to k_{y}'} \left ( k_{y}' - k_{y} \right )^2 \mbox{,}
\end{equation}
where $\nu_{N}= \left \{ \exp \left [ \left ( E_{N} - E_{\mathrm{F}} \right ) /k_{\mathrm{B}} T \right ] +1 \right \} ^{-1}$ is the LL filling factor, $E_{N}$ is the LL energy defined by Eq.\,(\ref{LandauLevels}), $E_{\mathrm{F}}$ is the Fermi energy, $k_{y}$ and $k_{y}'$ are the $y$-components of the electron wavevector before and after scattering, $D$ is the diffusion coefficient, and $W_{k_{y} \to k_{y}'}$ is the probability of scattering.

Two-phonon scattering in graphene in the QHR is possible through two different virtual intermediate states: a phonon with wavevector $\mathbf{q}^{+}$ is first emitted or a phonon with wavevector $\mathbf{q}^{-}$ is first absorbed. Transitions changing the electron LL number in the intermediate states are strongly suppressed due to small values of corresponding matrix elements and the presence of large denominators in the expression for $W_{k_{y} \to k_{y}'}$ (see Appendix). Therefore, in what follows we consider only transitions with no change in the electron LL number. Then, the probability of two-phonon scattering is given by Fermi's golden rule with the matrix element
\begin{equation}
\label{M_element_1}
M_{k_{y}, k_{y}'} \left(\mathbf{q}^{-} , \mathbf{q}^{+} \right) = n_{q^{-}} \left( n_{q^{+}} +1 \right) \times \sum_{k_{y}''} \left ( \frac {\langle k_{y}' \left | \hat{V}_{\mathbf{q}^{+}} \right | k_{y}'' \rangle \langle k_{y}'' \left | \hat{V}_{\mathbf{q}^{-}} \right | k_{y} \rangle} {\hbar \Omega_{q^{-}} } -  \frac {\langle k_{y}' \left | \hat{V}_{\mathbf{q}^{-}} \right | k_{y}'' \rangle \langle k_{y}'' \left | \hat{V}_{\mathbf{q}^{+}} \right | k_{y} \rangle}{\hbar \Omega_{q^{+}}} \right ) \mbox{,}
\end{equation}
where $ n_{q}=\left [ \exp \left ( \hbar \Omega_{q} / k_{\mathrm{B}} T \right ) - 1 \right ] ^{-1}$ is the phonon occupation number, $\Omega_{q}=sq$ is the acoustic phonon frequency, and $\hat{V}_{\mathbf{q}}$ is the electron-phonon coupling operator.

Operators describing electron scattering on intrinsic longitudinal (LA) and transverse (TA) acoustic phonons in graphene are \cite{Suzuura2002, Mariani2010}
\begin{equation}
\label{EFinteractionLA}
\hat{V}^{LA}_{\mathbf{q}} = U_{q} q \exp \left( i \mathbf{q} \mathbf{r} \right)
\begin{pmatrix}
i g_{d} & g_{h} e^{i 2 \varphi} \\ - g_{h} e^{- i 2 \varphi} & i g_{d}
\end{pmatrix}
\mbox{,}
\end{equation}

\begin{equation}
\label{EFinteractionTA}
\hat{V}^{TA}_{\mathbf{q}} = U_{q} q \exp \left( i \mathbf{q} \mathbf{r} \right)
\begin{pmatrix}
0 & i g_{h} e^{i 2 \varphi} \\ i g_{h} e^{- i 2 \varphi} & 0
\end{pmatrix}
\mbox{,}
\end{equation}
where $U_{q} = \left ( L_{x} L_{y} \right ) ^{-1/2} \left ( \hbar / 2 \rho \Omega_{q} \right ) ^{1/2}$, $\rho$ is the graphene 2D mass density, $\mathbf{r}$ is the position vector in the graphene plane, and $\varphi$ is the angle between the phonon wavevector $\mathbf{q}$ and the $x$-axis. The diagonal matrix elements in  Eqs.\,(\ref{EFinteractionLA})-(\ref{EFinteractionTA}) describe electron coupling to the phonon-created deformation potential, and the off-diagonal matrix elements originate from the phonon-induced bond-length modulations, which effect hopping amplitudes between two neighbouring sites. The corresponding coupling constants were estimated as $g_{d}\approx 20-30~\mathrm{eV}$ and $g_{h}\approx 1.5-3.0~\mathrm{eV}$.\cite{Ono1983, Klemens1994, Suzuura2002, Mariani2010, Pennington2003, Chen2008, Bolotin2008} Note that the deformation potential couples electrons with LA phonons only. Furthermore, since the second component of the electron wave function defined by Eq.\,(\ref{2componentWF}) vanishes for the zero LL, electrons in this LL do not interact with TA phonons.

Substituting the electron-phonon scattering operators given by Eqs.\,(\ref{EFinteractionLA})-(\ref{EFinteractionTA}) into Eq.\,(\ref{M_element_1}) yields
\begin{multline}
\label{M_element_2}
M_{k_{y}, k_{y}'}^{\mu, \gamma} \left(\mathbf{q}^{-} , \mathbf{q}^{+} \right) = G_{\mu, \gamma}^{2} U_{q^{-}} q^{-} n_{q^{-}} U_{q^{+}} q^{+} \left( n_{q^{+}} +1 \right) \\ \times \left ( \sum_{k_{y}''} \frac { M^{\mu, \gamma}_{ k_{y}, k_{y}''} \left(\mathbf{q^{-}}\right) M^{\mu, \gamma}_{ k_{y}'', k_{y}'} \left(\mathbf{q^{+}}\right) } { \hbar \Omega_{q^{-}}} - \sum_{k_{y}''} \frac { M^{\mu, \gamma}_{ k_{y}, k_{y}''} \left(\mathbf{q^{+}}\right) M^{\mu, \gamma}_{ k_{y}'', k_{y}'} \left(\mathbf{q^{-}}\right) } { \hbar \Omega_{q^{+}}} \right ) \mbox{,}
\end{multline}
where $\mu = \{LA,TA \}$, $\gamma = \{d, h\}$ are the indices introduced to separate the contributions to the longitudinal conductivity $\sigma_{xx}$ from LA and TA phonons and from the two different scattering mechanisms discussed above, $G_{\mu, \gamma}$ is the generalized electron-phonon coupling constant with the values $G_{LA, d}=g_{d}$, $G_{LA, h}=g_{h}$, $G_{TA, d}=0$, $G_{TA, h}=g_{h}$, and the matrix elements $M^{\mu, \gamma}_{ k_{y}, k_{y}'} \left(\mathbf{q}\right)$ are given by
\begin{equation}
\label{M_LA_d}
M^{LA,d}_{ k_{y}, k_{y}'} \left( \mathbf{q} \right) = i \int \Psi_{N, k_{y}'}^{\dagger} \left ( \mathbf{r} \right ) \exp \left(\pm i \mathbf{q} \mathbf{r}\right) I \Psi_{N, k_{y}} \left ( \mathbf{r} \right ) d \mathbf{r} = i C_{N}^{2} \exp \left(\pm \beta - \alpha\right) \left [ L_{N}^{0} \left ( \alpha \right ) + L_{N-1}^{0} \left ( \alpha \right ) \right ] \delta_{k_{y}',k_{y} \pm q_{y}} \mbox{,}
\end{equation}

\begin{multline}
\label{M_LA_h}
M^{LA,h}_{ k_{y}, k_{y}'} \left( \mathbf{q} \right) = \int \Psi_{N, k_{y}'}^{\dagger} \left ( \mathbf{r} \right ) \exp \left(\pm  i \mathbf{q} \mathbf{r} \right) \Phi_{LA} \Psi_{N, k_{y}} \left ( \mathbf{r} \right ) d \mathbf{r}  \\ = -  C_{N}^{2} N^{-1/2} l_{B} \left[ q_{x} \cos{2 \varphi} \pm \left( k_{y}-k_{y}' \right) \sin{2 \varphi} \right] \exp \left(\pm \beta - \alpha\right) L_{N-1}^{1} \left ( \alpha \right ) \delta_{k_{y}', k_{y} \pm q_{y}} \mbox{,}
\end{multline}

\begin{multline}
\label{M_TA_h}
M^{TA,h}_{ k_{y}, k_{y}'} \left( \mathbf{q} \right) = \int \Psi_{N, k_{y}'}^{\dagger} \left ( \mathbf{r} \right ) \exp \left(\pm  i \mathbf{q} \mathbf{r}\right) \Phi_{TA} \Psi_{N, k_{y}} \left ( \mathbf{r} \right ) d \mathbf{r}  \\ = - C_{N}^{2} N^{-1/2} l_{B} \left[ q_{x} \sin{2 \varphi} \mp \left( k_{y}-k_{y}' \right) \cos{2 \varphi} \right] \exp \left(\pm \beta - \alpha\right) L_{N-1}^{1} \left ( \alpha \right ) \delta_{k_{y}', k_{y} \pm q_{y}} \mbox{.}
\end{multline}
In Eqs.\,(\ref{M_LA_d})-(\ref{M_TA_h}), $\alpha = \left ( l_{B}^{2} / 4 \right ) \left [ \left ( k_{y}' - k_{y} \right )^2 + q_{x}^{2} \right ]$, $\beta = i \left ( l_{B}^2 q_{x} / 2 \right ) \left ( k_{y}' + k_{y} \right )$, $N \ne 0$, the ``$\pm$'' sign refers to emitted and absorbed phonons, $L_{N}^{0}$ and $L_{N}^{1}$ are the Laguerre polynomials, $I$ is the $2 \times 2$ identity matrix, and $\Phi_{LA}$, $\Phi_{TA}$ are given by
\begin{equation}
\Phi_{LA} =
\begin{pmatrix}
0 & e^{i 2 \varphi} \\ - e^{- i 2 \varphi} & 0
\end{pmatrix}
\mbox{, }
\Phi_{TA} =
\begin{pmatrix}
0 & i e^{i 2 \varphi} \\ i e^{- i 2 \varphi} & 0
\end{pmatrix}
\mbox{.}
\end{equation}
Eq.\,(\ref{M_LA_d}) is also valid for the zero LL when substituting $L^{0}_{N-1}=0$ and $L^{1}_{N-1}=0$. Note the $N^{-1/2}$ factor in Eqs.\,(\ref{M_LA_h})-(\ref{M_TA_h}) which suppresses the off-diagonal contribution to $\sigma_{xx}$ in higher LLs. Summation over all possible values of $k_{y}''$ in Eq.\,(\ref{M_element_2}) results in the following expressions for two-phonon scattering matrix elements
\begin{multline}
\label{M_LA_d_SQ}
\left| M_{k_{y}, k_{y}'}^{LA,d} \left(\mathbf{q}^{-} , \mathbf{q}^{+} \right) \right|^{2} = C_{N}^{8} \left( g_{d} U_{q} \right)^{4}  \left[ q  / \left( \hbar s \right) \right]^{2} \left[ L_{N}^{0} \left ( l_{B}^{2} q^{2}  / 2 \right ) + L_{N-1}^{0} \left ( l_{B}^{2} q^{2} / 2 \right ) \right]^{4} \\ \times \exp \left ( - l_{B}^{2} q^{2} \right ) \sin ^{2} \left [ l_{B}^{2} \left ( q_{y}^{+} q_{x}^{-} - q_{y}^{-} q_{x}^{+} \right ) / 2 \right ] \delta_{k_{y}', k_{y} + q_{y}^{-} - q_{y}^{+}} \mbox{,}
\end{multline}
\begin{multline}
\label{M_LA_h_SQ}
\left| M_{k_{y}, k_{y}'}^{LA,h} \left(\mathbf{q}^{-} , \mathbf{q}^{+} \right) \right|^{2} = C_{N}^{8}  \left( g_{h} U_{q} \right)^{4}  \left[ q  / \left( \hbar s \right) \right]^{2} N^{-2} \left[ L_{N-1}^{1} \left( l_{B}^{2} q^{2} / 2 \right) \right]^{4} \exp \left ( - l_{B}^{2} q^{2} \right ) \\ \times l_{B}^{4} \left(  q_{x}^{+} \sin{2 \varphi^{+}} + q_{y}^{+} \cos{2 \varphi^{+}} \right)^{2} \left(  q_{x}^{-} \sin{2 \varphi^{-}} + q_{y}^{-} \cos{2 \varphi^{-}} \right)^{2} \\ \times \sin ^{2} \left [ l_{B}^{2} \left ( q_{y}^{+} q_{x}^{-} - q_{y}^{-} q_{x}^{+} \right ) / 2 \right ] \delta_{k_{y}', k_{y} + q_{y}^{-} - q_{y}^{+}} \mbox{,}
\end{multline}
\begin{multline}
\label{M_TA_h_SQ}
\left| M_{k_{y}, k_{y}'}^{TA,h} \left(\mathbf{q}^{-} , \mathbf{q}^{+} \right) \right|^{2} = C_{N}^{8}  \left( g_{h} U_{q} \right)^{4}  \left[ q  / \left( \hbar s \right) \right]^{2} N^{-2} \left[ L_{N-1}^{1} \left( l_{B}^{2} q^{2} / 2 \right) \right]^{4} \exp \left ( - l_{B}^{2} q^{2} \right ) \\ \times l_{B}^{4} \left(  q_{x}^{+} \cos{2 \varphi^{+}} - q_{y}^{+} \sin{2 \varphi^{+}} \right)^{2} \left(  q_{x}^{-} \cos{2 \varphi^{-}} - q_{y}^{-} \sin{2 \varphi^{-}} \right)^{2} \\ \times \sin ^{2} \left [ l_{B}^{2} \left ( q_{y}^{+} q_{x}^{-} - q_{y}^{-} q_{x}^{+} \right ) / 2 \right ] \delta_{k_{y}', k_{y} + q_{y}^{-} - q_{y}^{+}} \mbox{.}
\end{multline}

Substituting the calculated probability of the two-phonon scattering $W_{k_{y} \to k_{y}' }$ into Eq.\,(\ref{DiffCoefficient}) and performing summation over $k_{y}'$ as well as integration over all possible values of $\mathbf{q}^{+}$ and $\mathbf{q}^{-}$ yield the following result for the longitudinal conductivity
$$
\sigma_{xx}= \left( \tilde {\sigma}_{xx}^{LA,d} + \tilde {\sigma}_{xx}^{LA,h} + \tilde {\sigma}_{xx}^{TA,h} \right)  \nu_{N} \left ( 1-\nu_{N} \right ) \mbox{,}
$$
where
\begin{multline}
\label{Sigma_d}
\tilde {\sigma}_{xx}^{LA,d}= \left( e^{2} / h \right) \left( C_{N}^{8} / 2\pi \right) \left(  g^{4}_{d} l_{B}  / \rho^{2} s^{4} \right) \left(T_{l_{B}} / T \right)  \\ \times \int  \limits_{0}^{\infty} \eta_{q} \left ( \eta_{q} +1 \right ) q^{4} \exp \left( - l_{B}^{2} q^{2} \right) \left [ 1 - J_{0} \left (l_{B}^{2} q^{2} \right ) \right ] \left [ L_{N}^{0} \left (l_{B}^{2} q^{2}/2 \right ) + L_{N-1}^{0} \left (l_{B}^{2} q^{2} / 2 \right ) \right ] ^{4} dq \mbox{,}
\end{multline}

\begin{multline}
\label{Sigma_h}
\tilde {\sigma}_{xx}^{LA/TA,h}= \left( e^{2} / h \right) N^{-2} \left( C_{N}^{8} / 2\pi \right) \left( g^{4}_{h} l_{B} / \rho^{2} s^{4} \right) \left( T_{l_{B}} / T \right) \\ \times \int \limits_{0}^{\infty} \eta_{q} \left ( \eta_{q} +1 \right ) q^{8} \exp \left(- l_{B}^{2} q^{2} \right) \left [ 1 - J_{0} \left (l_{B}^{2} q^{2} \right ) \right ] \left [L^{1}_{N-1} \left ( l_{B}^{2} q^{2} / 2 \right ) \right ] ^{4} dq \mbox{.}
\end{multline}
Here $J_{0}$ is the Bessel function of the first kind. The expressions for $\sigma_{xx}^{\mu, \gamma}$ were multiplied by a factor of four to account for the valley and spin degeneracy. From Eqs.\,(\ref{Sigma_d})-(\ref{Sigma_h}) it is evident that $\tilde {\sigma}_{xx}^{LA,d} \gg  \tilde {\sigma}_{xx}^{LA/TA,h}$ due to the relatively small value of $g_{h}$ comparing to $g_{d}$ \cite{Ono1983, Klemens1994, Suzuura2002, Mariani2010, Pennington2003, Chen2008, Bolotin2008} and the small $N^{-2}$ factor contained in Eq.\,(\ref{Sigma_h}).

There are two distinctive temperature limits. In the low-temperature limit, $T \ll T_{l_{B}}$, the main value of the integrals in Eqs.\,(\ref{Sigma_d})-(\ref{Sigma_h}) is formed by $q \le k_{B} T / \hbar s$ and $q l_{B}$ can be considered as a small parameter. Expanding the integrand into power series in $q l_{B}$ and taking into account only the lowest power term yields the following expression for $\sigma_{xx}^{\mu, \gamma}$ in the low-temperature limit
\begin{equation}
\label{Sigma_d_LT}
\tilde {\sigma}_{xx}^{LA,d} \simeq \left( e^2/h \right) \left( 2^{3} \pi ^{7} / 15 \right) \left[ {g}_{d}^{4} / \left( l_{B}^{2} \rho^{2} \hbar^{2} s^{6} \right) \right]  \left ( T/ T_{l_{B}} \right) ^{8} \mbox{,}
\end{equation}
\begin{equation}
\label{Sigma_h_LT}
\tilde {\sigma}_{xx}^{LA/TA,h} \simeq \left( e^2/h \right) \left[ A_{0}^{N-1} / N^{2} \right] \left( 5528 \pi ^{11}/ 1365 \right) \left[ {g}_{h}^{4} / \left( l_{B}^{2} \rho^{2} \hbar^{2} s^{6} \right) \right]  \left ( T/ T_{l_{B}} \right) ^{12} \mbox{.}
\end{equation}
In Eq.\,(\ref{Sigma_h_LT}) the coefficients $A_{0}^{N-1}$ are defined by $\left[ L_{N-1}^{1} \left( x/2 \right) \right]^{4}= \sum \limits_{j=0}^{4N-4} A_{j}^{N-1}  x^{j}$, where $N \ge 1$. For the zeroth LL, $\tilde {\sigma}_{xx}^{LA/TA,h}=0$. Note that $\tilde {\sigma}_{xx}^{LA,d}$ in Eq.\,(\ref{Sigma_d_LT}) does not depend on the LL number. The temperature dependencies given by Eqs.\,(\ref{Sigma_d_LT},\ref{Sigma_h_LT}) are different from the case of conventional semiconductor heterostructures for both two-phonon scattering \cite{Golovach2006, Bychkov1981, Dmitriev1995} and phonon-assisted hopping conductivity. \cite{Polyakov1994, Polyakov1995} In Eq.\,(\ref{Sigma_d_LT}), which corresponds to the deformation potential scattering mechanism, the lower power in the temperature dependence of mobility compared to that obtained in Refs.\,[\onlinecite{Golovach2006, Bychkov1981, Dmitriev1995}] is due to phonons in graphene being two-dimensional. As was mentioned above, the contribution to longitudinal conductivity at higher LLs given by Eq.\,(\ref{Sigma_h_LT}) is a distinctive feature of graphene without analogy in semiconductor systems.

In the more interesting high-temperature limit, $T > T_{l_{B}}$, the main value of the integrals in Eqs.\,(\ref{Sigma_d})-(\ref{Sigma_h}) is formed by $q \le 1/l_{B}$, and $\hbar s q / k_{\mathrm{B}} T$ can be considered as a small parameter. By expanding the integrand into power series in $\hbar s q / k_{\mathrm{B}} T$ and taking into account only the lowest power term we obtain the following result for $\sigma_{xx}^{\mu, \gamma}$ in the high-temperature limit
\begin{equation}
\label{Sigma_d_HT}
\tilde {\sigma}_{xx}^{LA,d}= \left( e^2/h \right) \left( C_{N} ^{8} \Phi_{N}^{d} / 2^{2} \pi \right) \left( {g}_{d}^{4} / l_{B}^{2} \rho^{2} \hbar^{2} s^{6} \right) \left ( T / T_{l_{B}} \right) \mbox{,}
\end{equation}
where
\begin{equation}
\label{Phi_d}
\Phi_{N}^{d} =\sum_{j=0}^{4N} B_{j}^{N} \Gamma \left ( j+ \frac {3} {2} \right ) \left [ 1- {}_{2}F_{1} \left ( \frac {2j+3} {4} ; \frac {2j+5} {4}; 1; -1 \right) \right ]\mbox{,}
\end{equation}
and
\begin{equation}
\label{Sigma_h_HT}
\tilde {\sigma}_{xx}^{LA/TA,h}= \left( e^2/h \right) N^{-2} \left[ \Phi_{N-1}^{h} / \left(2^{6} \pi \right) \right] \left[ {g}_{h}^{4} / \left( l_{B}^{2} \rho^{2} \hbar^{2} s^{6} \right) \right] \left ( T / T_{l_{B}} \right) \mbox{,}
\end{equation}
where
\begin{equation}
\label{Phi_h}
\Phi_{N-1}^{h}=\sum_{j=0}^{4N-4} A_{j}^{N-1} \Gamma \left ( j+ \frac {7} {2} \right ) \left [ 1-  {}_{2}F_{1} \left ( \frac {2j+7} {4} ; \frac {2j+9} {4}; 1; -1 \right) \right ] \mbox{.}
\end{equation}
In Eqs.\,(\ref{Phi_d})-(\ref{Phi_h}), $\Gamma$ is the Gamma function and ${}_{2}F_{1}$ is the hypergeometric function. In Eq.\,(\ref{Sigma_h_HT}), $N \ne 0$, since in the zeroth LL $\tilde {\sigma}_{xx}^{LA/TA,h}=0$. The coefficients $A_{j}^{N-1}$ are the same as in Eq.\,(\ref{Sigma_h_LT}) and the coefficients  $B_{j}^{N}$ are defined by $\left[ L_{N}^{0}\left( x/2 \right) +  L_{N-1}^{0}\left( x/2 \right) \right]^{4}= \sum \limits_{j=0}^{4N} B_{j}^{N}  x^{j}$.

\section{Discussion and conclusion}

\label{Discussion}

\begin{figure}[h]
\centering
\includegraphics*[width=0.49\linewidth]{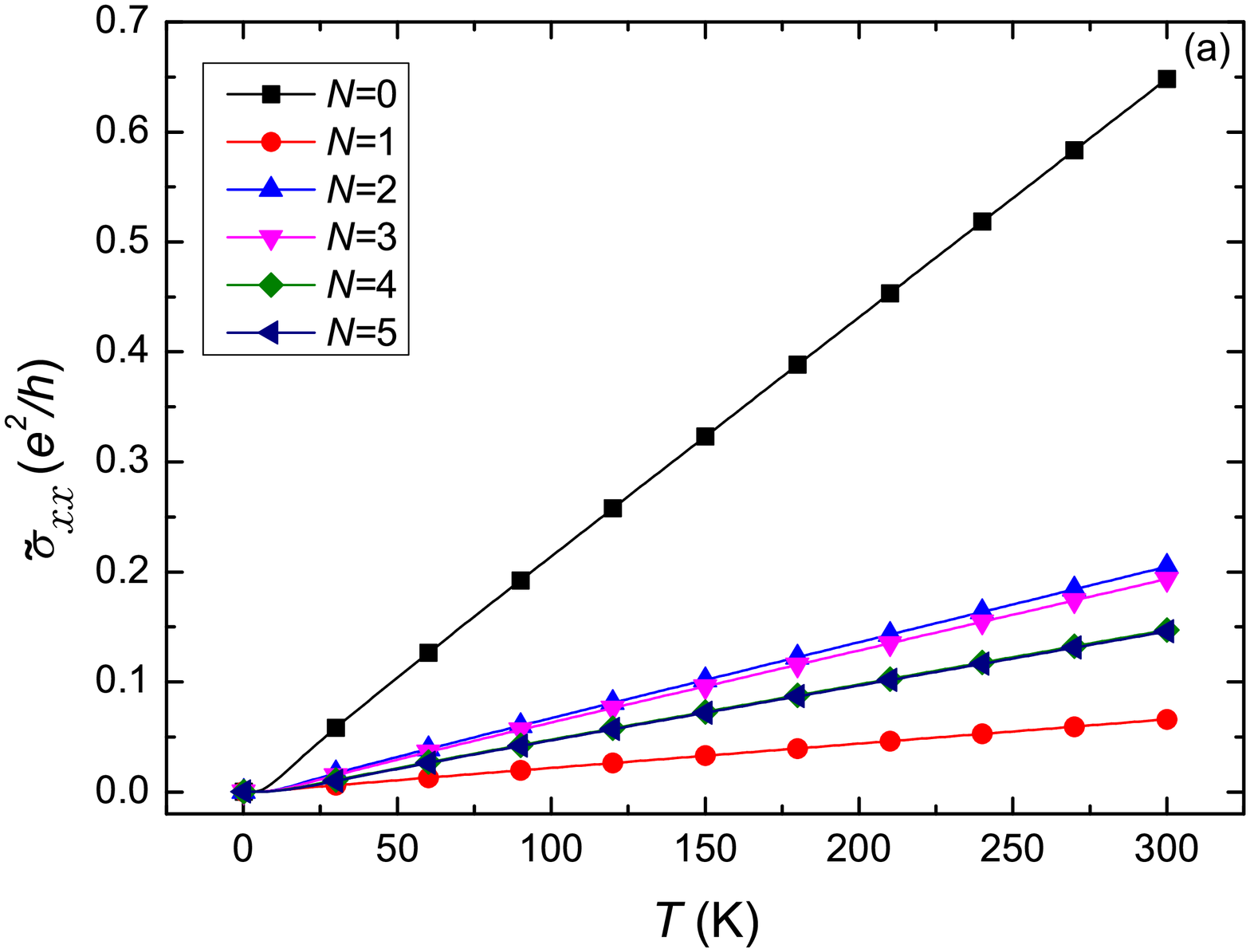}
\includegraphics*[width=0.49\linewidth]{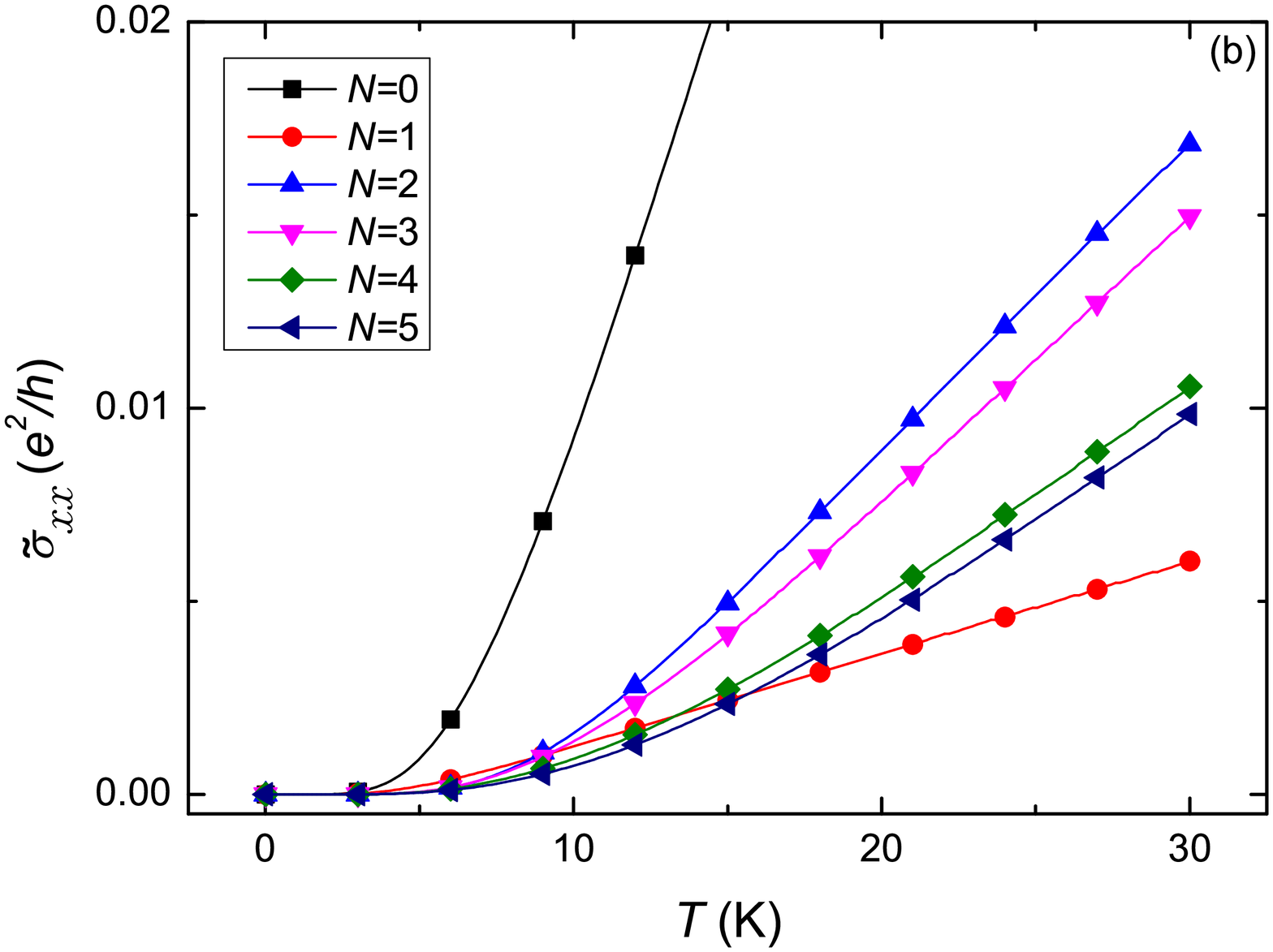}
\caption{(Colour online) (a) The longitudinal conductivity prefactor $\tilde {\sigma}_{xx}$ as a function of temperature for $B=10 \mbox{ } \mathrm{T}$. (b) Expanded view in the  $0-30 \mbox{ } \mathrm{ K}$ temperature range.}
\label{fig_sigma}
\end{figure}

In Fig.\,\ref{fig_sigma} we plot $\tilde {\sigma}_{xx}=\tilde {\sigma}_{xx}^{LA,d} + \tilde {\sigma}_{xx}^{LA,h} + \tilde {\sigma}_{xx}^{TA,h}$ obtained by numerical integration of Eqs.\,(\ref{Sigma_d})-(\ref{Sigma_h}) from $0$ to $300 \mathrm{K}$ for $B=10~\mathrm{T}$ ($T_{l_{B}} \approx 22 \mbox{ K}$).
It can be seen from Fig.\,\ref{fig_sigma}(a) that for $T \gg T_{l_{B}}$, $\tilde {\sigma}_{xx }$ linearly increases with temperature in line with Eqs.\,(\ref{Sigma_d_HT})-(\ref{Sigma_h_HT}). From Fig.\,\ref{fig_sigma}(b) one can see the change of the $\tilde {\sigma}_{xx}$ temperature dependence from the high-power to the linear law occurring around $T=T_{l_{B}}$.  Substituting the numerical values of all the constants into Eqs.\,(\ref{Sigma_d_HT}) and (\ref{Sigma_h_HT}) results in the following simplified expression for $\tilde {\sigma}_{xx}$ at elevated temperatures
\begin{equation}
\label{sigma_linearT}
\tilde {\sigma}_{xx} \approx \tilde {\sigma}_{N}  \left( T / 300 \mbox{ K} \right) \left( B / 10 \mbox{ T} \right)^{1/2} \mbox{,}
\end{equation}
where $\tilde {\sigma}_{N}$ has the following values for the six lowest LLs: $\tilde {\sigma}_{0}=0.65 e^{2}/h$, $\tilde {\sigma}_{1}=0.06 e^{2}/h$, $\tilde {\sigma}_{2}=0.20 e^{2}/h$, $\tilde {\sigma}_{3}=0.19 e^{2}/h$, $\tilde {\sigma}_{4}=0.15 e^{2}/h$, $\tilde {\sigma}_{5}=0.14 e^{2}/h$. Here we used the unscreened value of the deformation potential, which is arguably appropriate in the strong magnetic field as the electron motion is quantized and the screening is suppressed.

There is a natural question of the applicability of the lowest allowed (in our case second) order of perturbation theory in electron-phonon interaction at elevated temperatures. This problem was studied in detail in Ref.\,[\onlinecite{Dmitriev1995}] for conventional semiconductor quantum wells and bulk acoustic phonon scattering. To compare our results with the analysis provided in Ref.\,[\onlinecite{Dmitriev1995}] it is necessary to express them in terms of the diffusion coefficient
$$
D=\tilde {\sigma}_{xx} \frac {2 \pi l_{B}^2}{e^2}{k_{\mathrm{B}} T}=\frac{{\sigma}_{xx}}{e^2/h} \left(\frac{T}{T_{l_B}}\right) s l_B \mbox{.}
$$
In Ref.\,[\onlinecite{Dmitriev1995}] the diffusion coefficient for $T > T_{l_{B}}$ in the lowest (second) order of perturbation theory is written as $D=\left( \alpha T / T_{l_{B}} \right)^{2}s l_B$. Thus, there is a simple connection between the dimensionless electron-phonon interaction constant $\alpha$ and our dimensionless constants $\tilde {\sigma}_{N} / \left (e^{2} / h \right)$. Namely,
$$
\alpha=0.27 \left( \frac {\tilde \sigma_{N}} {4 e^{2} / h} \right)^{1/2} \left( \frac {B} {10 \mbox{ T}} \right)^{1/2} \mbox{.}
$$
According to Ref.\,[\onlinecite{Dmitriev1995}] higher orders of perturbation theory can be neglected for $T < T_{c}=T_{l_{B}} / \alpha$. Unlike the case of conventional semiconductors, for which $\alpha \propto l_{B}^{-2}$, for graphene $\alpha \propto l_{B}^{-1}$; therefore, the temperature $T_{c}$ defining the validity of perturbation theory becomes independent of magnetic field. For the zero Landau level and for the parameters used in our calculations, $T_{c} \approx 220 \mbox{ K}$. Above this temperature the phonon-induced mobility is expected to change from linear to sub-linear temperature dependence and eventually to saturate. For higher Landau levels the perturbation theory cut-off temperature $T_{c}$ well exceeds $300 \mbox{ K}$ and the lowest order perturbation analysis is fully valid for ambient conditions.

In conclusion, we obtained the value of the longitudinal conductivity in graphene in the quantum Hall regime due to two-phonon scattering at elevated temperatures which is comparable to the disorder-induced longitudinal magneto-conductivity in conventional semiconductor heterostructures.\cite{Polyakov1994, Polyakov1995,Clark1988} The predicted distinctive temperature and magnetic field dependence of the phonon scattering contribution to the pre-exponential factor in $\sigma_{xx}$ given by Eq.~(\ref{sigma_linearT}) can be easily separated from the temperature- and field-independent contribution caused by disorder when analysing experimental data. This should allow the parameters of electron-phonon interaction in graphene to be extracted with enhanced accuracy.

\begin{acknowledgments}
We thank Charles Downing for a critical reading of the manuscript. A.M.A. is grateful to Daniil Alexeev for fruitful discussions. This work was supported by the EU H2020 RISE project CoExAN (Grant No. H2020-644076), EU FP7 ITN NOTEDEV (Grant No. FP7-607521), FP7 IRSES projects CANTOR (Grant No. FP7-612285), QOCaN (Grant No. FP7-316432), and InterNoM (Grant No. FP7-612624). R.R.H. acknowledges financial support from URCO (Grant No. 15 F U/S 1TAY13-1TAY14) and Research Links Travel Grant by the British Council Newton Fund.
\end{acknowledgments}

\appendix*
\section{Electron-phonon scattering involving different Landau levels}

The probability of two-phonon scattering through a virtual intermediate state is calculated using Fermi's golden rule as given below
$$
\label{P_of_scattering}
W_{k_{y} \to k_{y}'}= \frac {2 \pi}{\hbar} \left| M_{k_{y}, k_{y}'}^{N_{i}, N_{f}} \left(\mathbf{q}^{-} , \mathbf{q}^{+} \right) \right|^{2} n_{q^{-}} \left( n_{q^{+}} +1 \right) \Delta \left (E_{i}, E_{f} \right ) \mbox{,}
$$
where
\begin{multline}
\label{M_element_suppl}
M_{k_{y}, k_{y}'}^{N_{i}, N_{f}} \left(\mathbf{q}^{-} , \mathbf{q}^{+} \right) = \sum_{N_{m}, k_{y}''} \frac {\langle N_{f}, k_{y}' \left | \hat{V}_{\mathbf{q}^{+}} \right | k_{y}'', N_{m} \rangle \langle N_{m}, k_{y}'' \left | \hat{V}_{\mathbf{q}^{-}} \right | k_{y}, N_{i} \rangle} {E_{i} - E_{m} + \hbar \Omega_{q^{-}} } \\ +  \sum_{N_{m}, k_{y}''} \frac {\langle N_{f}, k_{y}' \left | \hat{V}_{\mathbf{q}^{-}} \right | k_{y}'', N_{m} \rangle \langle N_{m} k_{y}'' \left | \hat{V}_{\mathbf{q}^{+}} \right | k_{y}, N_{i} \rangle}{E_{i} - E_{m} - \hbar \Omega_{q^{+}}} \mbox{,}
\end{multline}
and
$$
\label{G_delta_function}
\Delta \left (E_{i},E_{f} \right )=\delta \left (E_{i}-E_{f}-\hbar \Omega_{q^{+}} + \hbar \Omega_{q^{-}} \right ) \mbox{.}
$$
Here $E_{i}$, $E_{m}$, and $E_{f}$ are the energies of the electron LLs in the initial, intermediate and final states.  Clearly, the transitions changing the electron LL number in the intermediate states are suppressed compared to the transitions conserving the LL number due to the presence of large denominators in Eq.~(\ref{M_element_suppl}).

Notably, single-phonon scattering on acoustic phonons with a change in the LL number is also suppressed because of the very small value of the corresponding matrix element
$$
\langle N_{f}\left | \hat{V}^{LA/TA}_{\mathbf{q}} \right | N_{i} \rangle  \sim \exp \left( - l_{B}^{2} q^{2} / 4 \right) \mbox{.}
$$
Due to energy conservation, $q = \frac {\sqrt{2} v_{\mathrm{F}}} {s l_{B}} \left( \sqrt{N_{f}} - \sqrt{N_{i}} \right)$, which results in the following estimate
\begin{equation}
\label{exponent}
\langle N_{f} \left | \hat{V}^{LA/TA}_{\mathbf{q}} \right | N_{i} \rangle  \sim  \exp \left [- \frac{v_{\mathrm{F}}^{2}} {2 s^{2}} \left( \sqrt{N_{f}} - \sqrt{N_{i}} \right)^{2} \right] \mbox{.}
\end{equation}
Interestingly, unlike the case of a conventional quasi-two-dimensional semiconductor system,\cite{Erukhimov1969, Korneev1977} the exponential factor in  Eq.~(\ref{exponent}) does not depend on the magnetic field. Instead, it has a non-trivial dependence on the difference between the LL numbers of the two involved levels and becomes non-vanishing for very high adjacent LLs. However, these high levels are not relevant for the high-temperature quantum Hall effect, which is the subject of our interest; whereas, for $N_{i}=0$ and $N_{f}=1$, $\langle N_{f}\left | \hat{V}^{LA/TA}_{\mathbf{q}} \right | N_{i} \rangle \sim \exp \left( - v_{\mathrm{F}}^{2} / 2 s^{2} \right) \approx \exp \left( - 50^{2}/2\right)$ leading to a complete suppression of inter-LL transitions.


\begin{thebibliography}{60}
\bibitem{Klitzing1980}K.~von~Klitzing, G.~Dorda, and M.~Pepper, Phys. Rev. Lett. \textbf{45}, 494 (1980).

\bibitem{Jeckelmann1997}B.~Jeckelmann, B.~Jeanneret, and A.D.~Inglis, Phys. Rev. B \textbf{55}, 13124 (1997).

\bibitem{Jeckelmann2001}B.~Jeckelmann, B.~Jeanneret, Rep. Prog. Phys. \textbf{64} 1603 (2001).

\bibitem{Novoselov2004} K.\,S.~Novoselov, A.\,K.~Geim, S.\,V.~Morozov, D.~Jiang, Y.~Zhang, S.\,V.~Dubonos, I.\,V.~Grigorieva, and A.\,A.~Firsov, Science \textbf{306}, 666 (2004).

\bibitem{Novoselov2005}K.\,S.~Novoselov, A.\,K.~Geim, S.\,V.~Morozov, D.~Jiang, M.\,I.~Katsnelson, I.\,V.~Grigorieva, S.\,V.~Dubonos, and A.\,A.~Firsov, Nature (London) \textbf{438}, 197 (2005).

\bibitem{Zhang2005}Y.~Zhang, Y.-W.~Tan, H.\,L.~Stormer, and P.~Kim, Nature (London) \textbf{438}, 201 (2005).

\bibitem{McClure1956}J.\,W.~McClure, Phys. Rev. \textbf{104}, 666 (1956).

\bibitem{Geim2007}A.\,K.~Geim and K.\,S.~Novoselov, Nat. Mater. \textbf{6}, 183 (2007).

\bibitem{Neto2009}A.\,H.~Castro~Neto, F.~Guinea, N.\,M.\,R.~Peres, K.\,S.~Novoselov and A.\,K. Geim, Rev. Mod. Phys. \textbf{81}, 109 (2009).

\bibitem{Novoselov2007}K.\,S.~Novoselov, Z.~Jiang, Y.~Zhang, S.\,V.~Morozov, H.\,L.~Stormer, U.~Zeitler, J.\,C.~Maan, G.\,S.~Boebinger, P.~Kim, and A.\,K.~Geim, Science \textbf{315}, 1379 (2007).

\bibitem{Titeika1935}V.\,S.~Titeika, Ann. Phys. Leipzig \textbf{22}, 129 (1935).

\bibitem{Giesbers2008}A.\,J.\,M.~Giesbers, G.~Rietveld, E.~Houtzager, U.~Zeitler, R.~Yang, K.\,S.~Novoselov, A.\,K.~Geim, and J.\,C.~Maan, Appl. Phys. Lett. \textbf{93}, 222109 (2008).

\bibitem{Tzalenchuk2010}A.~Tzalenchuk, S.~Lara-Avila, A.~Kalaboukhov, S.~Paolillo, M.~Syv\"aj\"arvi, R.~Yakimova, O.~Kazakova, T.\,J.\,B.\,M.~Janssen, V.~Falko, and  S.~Kubatkin, Nat. Nanotechnol. \textbf{5}, 186 (2010).

\bibitem{Poirier2010}W.~Poirier and F.~Schopfer, Nat. Nanotechnol. \textbf{5}, 171 (2010).

\bibitem{Webber2013}J.\,A.\,Alexander-Webber, A.\,M.\,R.\,Baker, T.\,J.\,B.\,M.\,Janssen, A.\,Tzalenchuk, S.\,Lara-Avila, S.\,Kubatkin, R.\,Yakimova, B.\,A.\,Piot, D.\,K.\,Maude, and R.\,J.\,Nicholas, Phys. Rev. Lett. \textbf{111}, 096601 (2013).

\bibitem{Polyakov1994}D.\,G.~Polyakov and B.\,I.~Shklovskii, Phys. Rev. Lett. \textbf{73}, 1150 (1994).

\bibitem{Polyakov1995}D.\,G.~Polyakov and B.\,I.~Shklovskii, Phys. Rev. Lett. \textbf{74}, 150 (1995).

\bibitem{Apalkov2002}V.\,M.~Apalkov and M.\,E.~Portnoi, Phys. Rev. B, \textbf{66}, 121303 (2002).

\bibitem{Apalkov2002TwoLevelPRB}V.\,M.~Apalkov and M.\,E.~Portnoi, Phys. Rev. B, \textbf{65}, 125310 (2002).

\bibitem{Apalkov2002TwoLevelPhE}V.\,M.~Apalkov and M.\,E.~Portnoi, Physica E \textbf{15}, 202 (2002).

\bibitem{Golovach2006}V.\,N.~Golovach and M.\,E.~Portnoi, Phys. Rev. B \textbf{74}, 085321 (2006).

\bibitem{Ono1983}S.~Ono and K.~Sugihara, J. Phys. Soc. Jpn. 21, 861 (1966); K.~Sugihara, Phys. Rev. B \textbf{28}, 2157 (1983).

\bibitem{Klemens1994}P.\,G.~Klemens and D.\,F.~Pedraza, Carbon \textbf{32}, 735 (1994).

\bibitem{Mariani2008}E.~Mariani and F.~von~Oppen, Phys. Rev. Lett. \textbf{100}, 076801 (2008).

\bibitem{Castro2010}E.\,V.~Castro, H.\,Ochoa, M.\,I.~Katsnelson, R.\,V.~Gorbachev, D.\,C.~~Elias, K.S.~Novoselov, A.\,K.~Geim, and F.~Guinea, Phys. Rev. Lett. \textbf{105}, 266601 (2010).

\bibitem{Wang2013}L.~Wang, I.~Meric, P.\,Y.~Huang, Q.~Gao, Y.~Gao, H.~Tran, T.~Taniguchi, K.~Watanabe, L.\,M.~Campos, D.\,A.~Muller, J.~Guo, P.~Kim, J.~Hone, K.\,L.~Shepard, and C.\,R.~Dean, Science \textbf{342}, 614 (2013).

\bibitem{Geick1966}R.~Geick and C.\,H.~Perry, Phys. Rev. \textbf{146}, 543 (1966).

\bibitem{Hwang1978}H.\,C.~Hwang and J.~Henkel, Phys. Rev. B \textbf{17}, 4100 (1978).

\bibitem{Lazzeri2006}M.~Lazzeri, S.~Piscanec, F.~Mauri, A.\,C.~Ferrari, and J.~Robertson, Phys. Rev. B \textbf{73}, 155426 (2006).

\bibitem{Suzuura2008}H.~Suzuura and T.~Ando, J. Phys. Soc. Jpn. \textbf{77}, 044703 (2008).

\bibitem{Goerbig2007}M.\,O.~Goerbig, J.-N.~Fuchs, K.~Kechedzhi, and V.\,I.~Fal’ko, Phys. Rev. Lett. \textbf{99}, 087402 (2007).

\bibitem{Mori2011}N.~Mori and T.~Ando, J. Phys. Soc. Jpn. \textbf{80}, 044706 (2011).

\bibitem{Kim2013}Y.~Kim, J.\,M.~Poumirol, A.~Lombardo, N.\,G.~Kalugin, T.~Georgiou, Y.\,J.~Kim, K.\,S.~Novoselov, A.\,C.~Ferrari, J.~Kono, O.~Kashuba, V.\,I.~Fal’ko, and D.~Smirnov, Phys. Rev. Lett. \textbf{110}, 227402 (2013).

\bibitem{Giesbers2007}A.\,J.\,M.~Giesbers, U.~Zeitler, M.\,I.~Katsnelson, L.\,A.~Ponomarenko, T.\,M.~Mohiuddin, and J.\,C.~Maan, Phys. Rev. Lett. \textbf{99}, 206803 (2007).

\bibitem{Pereira2011}A.\,L.\,C.~Pereira, C.\,H.~Lewenkopf, and E.\,R.~Mucciolo, Phys. Rev. B \textbf{84}, 165406 (2011).

\bibitem{Bychkov1981}Yu.\,A.~Bychkov, S.\,V.~Iordanski, and G.\,M.~Eliashberg, Pis’ma
Zh. Eksp. Teor. Fiz. \textbf{34}, 496 (1981) [JETP Lett. \textbf{34}, 473 (1981)].

\bibitem{Dmitriev1995} A.\,P.~Dmitriev and V.\,Yu.~Kacharovskii, Phys. Rev. B \textbf{52}, 5743 (1995).

\bibitem{Suzuura2002}H.~Suzuura and T.~Ando, Phys. Rev. B \textbf{65}, 235412 (2002).

\bibitem{Mariani2010}E.~Mariani and F.~von~Oppen, Phys. Rev. B \textbf{82} 195403 (2010).

\bibitem{Pennington2003}G.~Pennington and N.~Goldsman, Phys. Rev. B \textbf{68}, 045426 (2003).

\bibitem{Chen2008}J.\,H.~Chen, C.~Jang, S.~Xiao, M.~Ishigami, and M.\,S.~Fuhrer, Nat. Nanotechnol. \textbf{3}, 206 (2008).

\bibitem{Bolotin2008}K.\,I.~Bolotin, K.\,J.~Sikes, J.~Hone, H.\,L.~Stormer, and P.~Kim, Phys. Rev. Lett \textbf{101}, 096802 (2008).

\bibitem{Clark1988}R.\,G.~Clark, J.\,R.~Mallett, A.~Usher, A.\,M.~Suckling, R.\,J.~Nicholas, S.\,R.~Haynes, Y.~Journaux, J.\,J.~Harris, and C.\,T.~Foxon, Surface Science \textbf{196}, 219 (1988).

\bibitem{Erukhimov1969}M.\,Sh.~Erukhimov, Fiz. Tekh. Poluprovodn. \textbf{3}, 194 (1969) [Sov. Phys. Semicond. \textbf{3}, 162 (1969)].

\bibitem{Korneev1977}V.\,V.~Korneev, Fiz. Tverd. Tela (Leningrad) \textbf{19}, 357 (1977) [Sov. Phys. Solid State \textbf{19}, 205 (1977)].

\end{thebibliography}
\end{document}